\documentclass[twocolumn,showpacs,preprintnumbers,amsmath,superscriptaddress,amssymb,prl,aps,10pt]{revtex4-1}
\usepackage{graphicx}
\usepackage{dcolumn}
\usepackage{bm}
\usepackage{ulem}
\usepackage{color}
\usepackage{float}
\newcommand{\mr}{\mathrm}

\begin{document}

\title{Coherence properties of a 2D trapped Bose gas around the superfluid transition}

\author{T. Plisson}
\affiliation{Laboratoire Charles Fabry, Institut d'Optique, CNRS, Univ Paris-Sud, 2, Avenue Augustin Fresnel, 91127 PALAISEAU
CEDEX}
\author{B. Allard}
\affiliation{Laboratoire Charles Fabry, Institut d'Optique, CNRS, Univ Paris-Sud, 2, Avenue Augustin Fresnel, 91127 PALAISEAU
CEDEX}
\author{M. Holzmann}
\affiliation{LPTMC, UMR 7600 of CNRS, Universit\'e P. et M. Curie, 75752 Paris, France}
\affiliation{Universit\'e Grenoble 1/CNRS, LPMMC, UMR 5493, B.P. 166, 38042 Grenoble, France}
\author{G. Salomon}
\affiliation{Laboratoire Charles Fabry, Institut d'Optique, CNRS, Univ Paris-Sud, 2, Avenue Augustin Fresnel, 91127 PALAISEAU
CEDEX}
\author{A. Aspect}
\affiliation{Laboratoire Charles Fabry, Institut d'Optique, CNRS, Univ Paris-Sud, 2, Avenue Augustin Fresnel, 91127 PALAISEAU
CEDEX}
\author{P. Bouyer}
\affiliation{Laboratoire Charles Fabry, Institut d'Optique, CNRS, Univ Paris-Sud, 2, Avenue Augustin Fresnel, 91127 PALAISEAU
CEDEX}
\affiliation{LP2N, Univ Bordeaux 1, IOGS, CNRS, 351 cours de la Lib\'eration, 33405 Talence, France}
\author{T. Bourdel}
\email[Corresponding author: ]{thomas.bourdel@institutoptique.fr}
\affiliation{Laboratoire Charles Fabry, Institut d'Optique, CNRS, Univ Paris-Sud, 2, Avenue Augustin Fresnel, 91127 PALAISEAU
CEDEX}

\date{\today}
\begin{abstract}
We measure the momentum distribution of a 2D trapped Bose gas and observe the increase of the range of coherence around the Berezinskii-Kosterlitz-Thouless (BKT) transition. We quantitatively compare our observed profiles to both a Hartee-Fock mean-field theory and to quantum Monte-Carlo simulations. In the normal phase, we already observe a sharpening of the momentum distribution. This behavior is partially captured in a mean-field approach, in contrast to the physics of the BKT transition.
\end{abstract}


\maketitle
The characteristics of a Bose gas in the degenerate regime are
greatly dependant on the dimensionality of the system. In contrast
to its three-dimensional (3D) counterpart, a two-dimensional
(2D) interacting Bose gas does not present true long range order
at low temperature \cite{Mermin66,Hohenberg67}. Instead, it
undergoes a superfluid-to-normal fluid transition described by the
BKT theory \cite{Kosterlitz73,Berezinski72}. This superfluid
transition has been observed in $^4\mr{He}$ films \cite{Bishop78},
Josephson coupled superconducting arrays \cite{Resnick81}, and
recently in dilute ultracold atomic gases \cite{Hadzibabic06}.

In ultracold 2D trapped Bose gases, the apparition of superfluidity 
was predicted to be preceded by the reduction of density fluctuations \cite{Prokofev01,Prokofev02} and by the increase of coherence \cite{Bisset09,Holzmann10}. Experimentally, great efforts were devoted to
the study of the density distribution $n(\bf{r})$ of 2D
gases, which gives direct access to the phase space density and to
the density fluctuations. It has permitted the study of the scale
invariance \cite{Rath10} and universality in 2D Bose gases \cite{Hung11} as
well as its thermodynamics \cite{Yefsah11}. The study of the first order correlation function $g_1$ is possible in real space through 
interferences between two clouds of atoms \cite{Hadzibabic06,Clade09} which revealed its expected algebraic decay in the superfluid regime \cite{Hadzibabic06}. 
The Fourier transform of $g_1$ is also naturally embedded  in the momentum distribution \cite{Gerbier03, Richard03}. For example, the narrow peak in the density distribution after time of flight is frequently used as indicator of the slow algebraic decay of the coherence in the superfluid regime  \cite{Kruger07, Clade09, Tung10}.

In this paper, we present a detailed
analysis of the experimental momentum distribution of
2D Bose gases closer to a genuine 2D gas than previous studies \cite{Tung10}. In addition, we fit our profiles using a Hartree-Fock mean-field (HFMF) model already used in density space \cite{Holzmann08} but extended here for the momentum distribution, and compare our results to quantum Monte-Carlo (QMC) simulations \cite{Holzmann08,Holzmann10} . As we increase the phase space density, we observe a progressive narrowing of the momentum distribution and a change of shape from an approximately gaussian to a peaked distribution. In particular, at the superfluid phase transition, the momentum distribution is already peaked and its width reduced by a factor $\sim5$ compared to a Boltzmann thermal gas. Whereas this behavior is partially captured in a mean-field approach where the 2D Bose statistics plays a central role, correlation effects beyond mean field are cleary visible in the full momentum profile.

\begin{figure}[t!]
\includegraphics[width=0.48\textwidth]{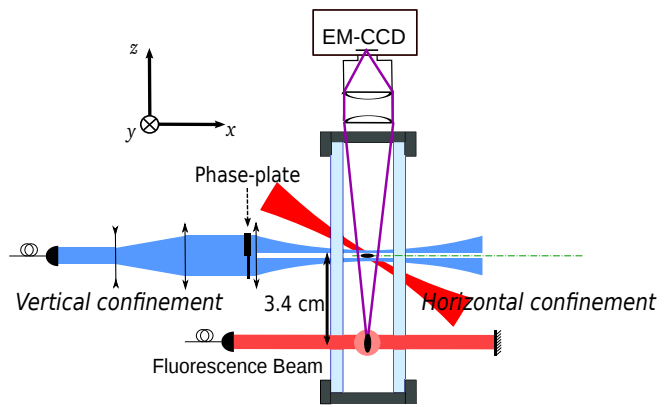}
\caption{Scheme of the experimental setup.The atom cloud, initially trapped in a harmonic 2D trap with frequencies 8\,Hz $\times$ 15\,Hz $\times$ 1.5\,kHz, is dropped for $83.5~\mr{ms}$ before shining a fluorescence beam situated $~3.4~\mr{cm}$ below the initial position. The fluorescence signal is recorded from above with an electron-multiplying-CCD camera.}
\label{fig1}
\end{figure}

Experimentally, the 2D Bose gases are prepared as follows. We
prepare a 3D cloud of $^{87}\mathrm{Rb}$ atoms cooled by
all-optical runaway evaporation in a crossed dipole trap, as
described in \cite{Clement09}. During the evaporation process, we
turn on a uniform magnetic gradient to select a single hyperfine
state ($|\mr{F=1}$, $\mr{m}_\mr{F}=0\rangle$) \cite{Couvert08}. We then transfer the cloud to a 2D trap. For that, we adiabatically ramp on a blue detuned laser at
$767~\mr{nm}$ in a $\mr{TEM}_{01}$-like mode, providing an intensity node in the horizontal plane
of the atoms (see Fig.\ref{fig1}). It induces a strong confinement
in the vertical direction, with an oscillation frequency of
$\omega_z/2\pi=1.5~\mr{kHz}$, corresponding to a
dimensionless interaction strength
$\tilde{g}=g m/\hbar^2=\sqrt{8\pi}a_\mr{s}/a_z=0.096$, where
$a_\mr{s}=5.3~\mr{nm}$ is the 3D scattering length,
$a_z=\sqrt{\hbar/m\omega_z}$ is the harmonic oscillator length
in the vertical direction, $\hbar$ is the reduced Planck constant, and $m$
is the atomic mass.

Horizontally, the 2D trap is made of a laser beam at
$1565~\mr{nm}$ with a waist $\mr{w}=200~\mu \mr{m}$ and tilted by
$\sim~30^\circ$ with respect to the horizontal plane (see Fig.\ref{fig1}).
Its oscillation frequencies are $\omega_x/2\pi=8~\mr{Hz}$,
$\omega_y/2\pi=15~\mr{Hz}$. The atom number $N$ is varied
from $2\times 10^4$ to $6\times 10^4$ by changing the number of
atoms initially loaded. The final temperature remains
approximately constant at $T=64.5~\mr \pm~2.0~\mr{nK}$. The two dimensional character of our experiment is given by
$k_\mr{B}T/\hbar\omega_z=0.90$ where $k_\textrm{B}$ is the Boltzmann constant, which results in having
$\sim~70~\%$ of the atoms in the ground state of the vertical
harmonic oscillator.

We let the atoms thermalize for $500~\mr{ms}$ in the final trap
before probing the momentum distribution function through time of
flight imaging (TOF). In the first milliseconds of expansion, the gas
expands predominantly in the vertical direction and the
interaction energy is thus quickly released in this direction. As
a result, the horizontal momentum distribution (in the $xy$ plane)
remains unchanged during the expansion. After an expansion time of
$t_\textrm{TOF}=83.5~\mr{ms}$ much longer than $1/\omega_x$
and $1/\omega_y$, the horizontal density distribution
reflects the initial momentum distribution \cite{footnote}. Finally, two circularly polarized saturating
retro-reflected laser beams, resonant with the $\left |
5\mr{S}_{1/2},~\mr{F}=2\right > \to \left |
5\mr{P}_{3/2},~\mr{F'}=3\right >$ transition and with the $\left |
5\mr{S}_{1/2},~\mr{F}=1\right > \to \left |
5\mr{P}_{3/2},~\mr{F'}=2\right >$ transition allow the atoms to
fluoresce for $100~\mu \mr{s}$. The fluorescence signal is
recorded on an electron-multiplying-CCD camera placed along the
vertical axis, thus imaging the horizontal profile.

\begin{figure}[t!]
\includegraphics[width=0.48\textwidth]{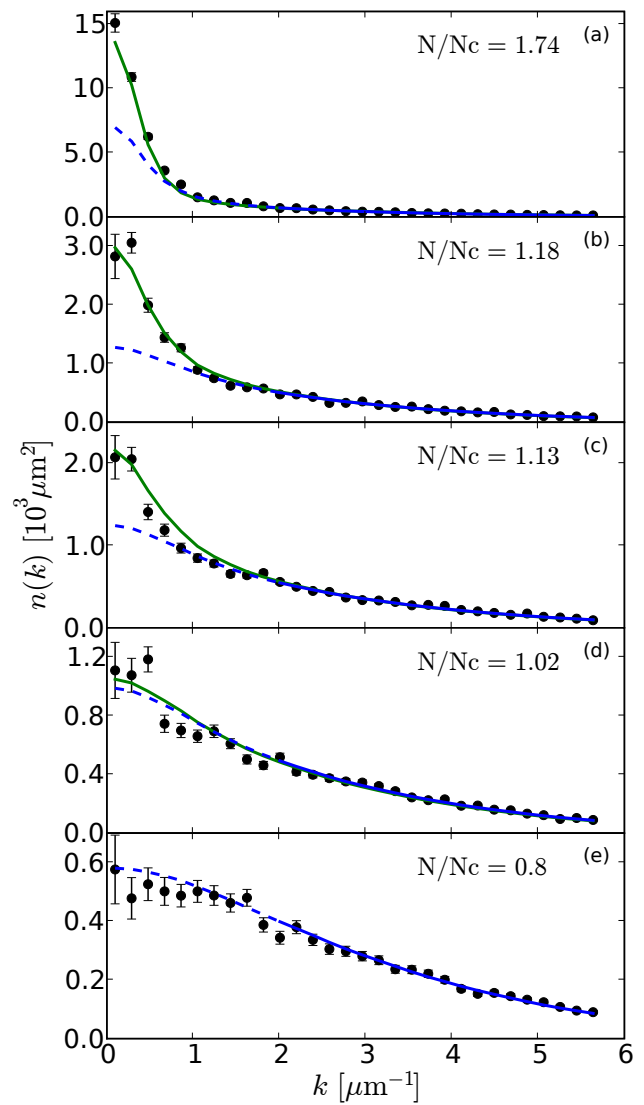}
\caption{Radial profile of the momentum distribution for five different atom numbers at constant temperature $T=64.5~\pm~2.0$ nK. The atom number $N$ is given in unit of the critical number for the ideal gas Bose-Einstein condensation $N_\mr{c}\approx 3\times 10^4$ and the superfluid phase transition is expected for $N/N_\mr{c}\approx 1.18$ (see text). In each plot, we present the
experimental data (black dots), the mean-field profiles (blue dashed line) resulting from the fit
in the wing (i.e. for $k>2\,\mu$m$^{-1}$) and the corresponding quantum
Monte-Carlo profiles (green continuous line) with the same temperature and the measured atom number.
} \label{fig2}
\end{figure}

To analyze the experimental data, we perform an azimuthal
averaging of the single pictures, as described in \cite{Rath10},
in order to extract radial profiles corresponding to the momentum
distributions $n(|{\bf k}|)$, where ${\bf k}$ is the atom wavevector. Experimentally, we
weight every point of the profile by its experimental standard
deviation (coming from the photon shot noise, the camera dark
noise and the number of averaging points). Such profiles are presented in
Fig.\,\ref{fig2}. For a low atom number (see Fig.\,\ref{fig2}e), the distribution
is approximately gaussian and relatively broad. For
a higher atom number (Fig.\,\ref{fig2}c-d), the curve progressively peaks and deviates from a gaussian. At high atom number (Fig.\,\ref{fig2}a-b), a sharp feature develops at very low momentum and grows with increasing atom number. Its width tends to a constant and agrees with our resolution in momentum space
(half width of $\sim~0.3 \, \mu$m$^{-1}$) coming essentially from
our imaging resolution and from the initial size of the cloud.  Since the peak at low momentum develops on a distribution that is
not a gaussian, it is hard to precisely point when it first appears.

In order to attribute a temperature to each profile, we have to
rely on a model. More precisely, we fit the Hartree-Fock mean-field model to the wings of the momentum
distribution data. This model has
already been used for in-situ density profiles \cite{Holzmann10, Rath10, Holzmann08} but needs to be extended to get the momentum distribution. We proceed as follows.
The density distribution in HFMF theory, in the local density
approximation, reads
\begin{equation}
         n({\bf r})=\frac{1}{(2\pi)^2}\int_0^{\infty} \frac{2\pi k \mr{d}k}{e^{\beta(\hbar^2 k^2/2m+2gn({\bf r})-\mu({\bf r}))}-1}~,
\end{equation}
 where $\beta=1/k_\textrm{B}T$,
$\mu({\bf r})=\mu_0-m\omega_x^2
x^2/2-m\omega_y^2 y^2/2$ is the local chemical potential with $\mu_0$ the chemical potential
at the trap center.
After integration, this leads to the following equation relating
$\mu({\bf r})$ and $ n({\bf r})$ \cite{Hadzibabic09}:
\begin{equation}
\beta \mu({\bf r}) = 2\beta gn({\bf r}) +\ln\left(1-e^{-\lambda_\mr{dB}^2 n({\bf r})}\right)
\label{mu_n}~,
\end{equation}
 where $\lambda_\mr{dB}=\sqrt{2 \pi \hbar^2/m k_\textrm{B}T}$ is
the thermal de Broglie wavelength. Writing
the momentum distribution:
\begin{equation}
n({\bf k})= \frac{1}{(2\pi)^2}\int \frac{\mr{d}x
\mr{d}y}{e^{\beta(\hbar^2
k^2/2m+2gn({\bf r})-\mu({\bf r}))}-1} \label{n_k1}
\end{equation}
and making the change in variables from $(x, y)$ to $\mu$, we
directly obtain $n({\bf k})$ after integration. By taking into account
the thermally populated vertical levels but neglecting the interaction in
these levels, which is justified since the densities in the
excited levels are small, we finally calculate the function\,:
\begin{equation}
\begin{aligned}
n({\bf k})&= \frac{1}{2\pi m \omega^2}\int_{-\infty}^{\mu_0} \frac{\mr{d}\mu}{e^{\beta(\hbar^2 k^2/2m+2gn(\mu)-\mu))}-1} \\
& + \sum_{\nu>0}\frac{-1}{2\pi\beta m \omega^2 }\ln\left(1-e^{-\beta(\hbar^2k^2/2m+\nu \hbar \omega_z -\mu_0)}\right)\\
\end{aligned}
\label{n_k}
\end{equation}
Where $ n(\mu)$ is determined numerically by solving Eq.(\ref{mu_n}), and where $\omega^2=\omega_x^2+\omega_y^2$.
For given values of $\mu_0$ and $T$, we can calculate $n(|{\bf 
k}|)$ and then use this function to fit to the experimental
data. 

More precisely, we fit the wings of the data where we
expect beyond mean-field effects to play little role and we can extract temperature and chemical potential. This requires an accurate calibration of the atom detection efficiency, which is a rather difficult task. We have performed QMC simulations based on numerically exact path-integral  algorithm \cite{Krauth96,Holzmann99,Holzmann08} to calculate the momentum profiles, so that  we can adjust the calibration of the atom detector. Within the experimental
error bars, it agrees with an independent calibration using the 3D
condensation threshold. Moreover, by fitting a mean-field model to the wings of QMC simulations, we find the accurate temperature and thus validate our fitting method.  

In Fig.\ref{fig2}, we plot with each profile the fitted mean-field distribution (blue dashed line). We observe that the mean-field fit reproduces well the experimental data at low atom number, while it fails accounting for the central part of the profiles when the number of atoms increases. However, the QMC simulations for the fitted temperatures and the experimental atom numbers (green continuous lines in Fig.\ref{fig2}) are in agreement with the experiment in all regimes. For both the calculated mean-field and QMC profiles, we take into account the finite resolution of our imaging system. 

In order to quantify the degeneracy of the gas, we calculate
$N/N_\mr{c}$, where
\begin{equation}
N_\mr{c}=\sum_{\nu}(\beta\hbar\omega)^{-2}g_2(e^{-\nu\beta\hbar\omega_z})
\end{equation}
is the critical atom number for a non-interacting Bose gas for our trap
parameters, and $g_2(x)=\sum_{n=1}^{\infty}x^n/n^2$.
From classical field calculations \cite{Prokofev01,Prokofev02,Holzmann07},
the BKT transition is expected at a central density of the lowest vertical level $\ln(380.3/\tilde{g})/\lambda_{\mr{dB}}^2$ with small quantum corrections of order $\tilde{g}/\lambda_{\mr{dB}}^2$ \cite{Holzmann08,Holzmann10}. Integrating the dominating classical field corrections to mean-field \cite{Holzmann10} yields $(N/N_\mr{c})_{\mr{BKT}} \approx 1.18$ as critical value of the particle number. Close to this value, the experimental profiles are already peaked at low momentum (Fig.\ref{fig2}b-c). 

\begin{figure}[t!]
   \includegraphics[width=0.48\textwidth]{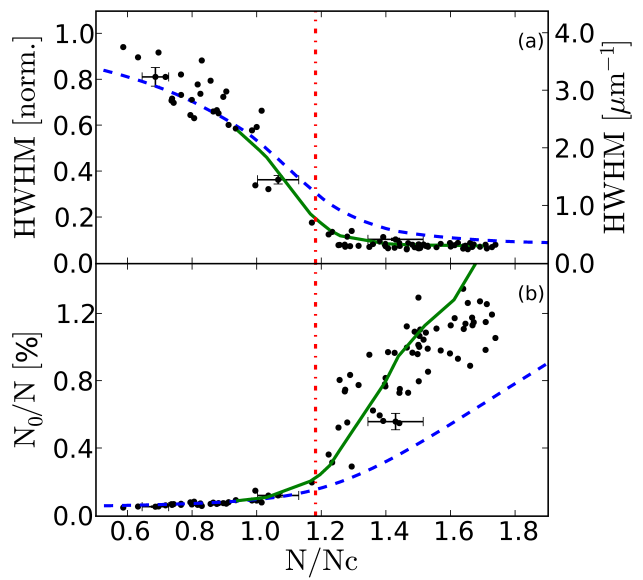}
  \caption{a. Half width at half maximum of the profiles,
normalized to the HWHM of a gaussian profile for the same
temperature, plotted versus the number of atoms normalized to
$N_\textrm{c}$. b. Fraction of atoms in the central pixel of the
image $N_0/N$. Blue dashed line : mean-field prediction for
$T=64.5~\mr{nK}$. Green continuous line: Monte-Carlo data for $T=64.5~\pm~0.3~\mr{nK}$. Red vertical dot-dashed line : superfluid transition. }
  \label{fig3}

\end{figure}

In order to analyze our experimental findings further, we now
consider two quantities : the half width at half maximum (HWHM)
and the fraction of atoms in the central pixel $N_0/N$. The advantage of those two quantities is that they are
model-independent and quantify the degree of coherence
of the gas. The HWHM gives a measure of the inverse of the
coherence length whereas the fraction of atoms in the central
pixel is related to the fraction of atoms which are coherent on a
length scale larger than $\sim~5~\mu\mr{m}$. In Fig.\ref{fig3}, we plot
these two quantities as a function of $N/N_\mr{c}$. The HWHM is
normalized to the one of a gaussian distribution : $2\sqrt{\pi\ln(2)}/\lambda_{\mr{dB}}=4.0~\mu\mr{m}^{-1}$. Already
for $N/N_\mr{c}\approx 0.5$, the normalized width of the momentum
distribution starts to decrease from 1. In other word, the
distribution is not a gaussian any more and tends to peak at low
momentum. At the superfluid transition point
($N/N_\mr{c}\approx 1.18)$, the HWHM has already decreased by a
factor of $\sim5$. After the transition, the HWHM saturates to a value corresponding to the imaging resolution.

The change of shape in the momentum distribution is also reflected
in the fraction of atoms in the central pixel (Fig. 3b). This
value increases smoothly but quite dramatically as the number of
atoms increases. For low atom number, it corresponds to the decrease of the width and for high atom number, while the HWHM saturates, the fraction of atoms in the central pixel keeps growing, reflecting the increasing coherence of the gas with atom number. It is
interesting to note that the superfluid transition does not appear
as a sharp feature in our data on the coherence properties of the
gas, unlike what is expected for the superfluid fraction \cite{Hadzibabic09, Bisset09}. 

In addition to the experimental points, we also plot Monte-Carlo
simulations and mean-field calculations for our experimental
conditions. The
Monte-Carlo simulations show a good agreement with our
experimental findings. The mean-field results coincides with the Monte-Carlo simulations at low atom number up to about
$N/N_\mr{c}\approx 1$. For high degeneracy parameters $N/N_\mr{c}
> 1$, the mean-field model underestimates the height of the coherence
peak showing that beyond mean-field effects become important.
It is remarkable that the mean-field approximation captures the
initial increase of the coherence length. This effect is thus not in direct link with the physics of the BKT phase transition
for which the theory is inherently beyond mean-field. In fact, even in a
non-interacting trapped Bose gas, for which calculations are exact,
there is also an increase of the coherence length before the
Bose-Einstein phase transition. 

The presence of a low momentum coherence peak cannot be considered as a signature of the BKT phase transition. An accurate signature however is the slow algebraic decay of the first order coherence function (as $1/r^\alpha$ with $\alpha<0.25$) \cite{Hadzibabic06}.  This decay can in principle be observed in the momentum distribution and indeed our QMC calculations show a change of slope (to negative) in the function $k^{2-\alpha}n(k)$ at low momentum,
$k < \lambda_\mr{dB}$ for $N \gtrsim 1.2N_c$, together with the onset of superfluidity. Unfortunately, our experimental resolution is not sufficient for direct observation of this feature.

In conclusion, we have studied the momentum distribution of a
trapped interacting 2D Bose gas. 
Our finding are in qualitative agreement with
previously published results \cite{Clade09, Tung10, Bisset09} but we analyze our profiles in more
details and compare them to mean-field and Monte-Carlo
calculations. In particular, we show that the momentum distribution narrows progressively and well before the BKT phase transition. Our detailed characterization of the 2D trapped Bose gas momentum distribution will be a useful tool for further studies of the remarkable properties of 2D gases of ultracold atoms. 

We thank M. Robert-de-Saint-Vincent for early contributions to the experiment, F. Moron and A. Villing for technical assistance, J.
Dalibard, T. Giamarchi, and W. Krauth for discussions. This
research was supported by CNRS, Ministere de l'Enseignement
Sup\'erieur et de la Recherche, Direction G\'en\'erale de
l'Armement, ANR-08-blan-0016-01, IXBLUE, RTRA: Triangle de la
physique, EuroQuasar program of the ESF and EU, iSense. LCFIO is
member of IFRAF.


\begin{thebibliography}{00}
\bibitem{footnote}
For the detailed analysis of the profiles, we slightly correct $t_\textrm{TOF}$ to the value $((t_\textrm{TOF}^2+1/\omega_x^2).(t_\textrm{TOF}^2+1/\omega_y^2))^{1/4}=85~\mr{ms}$, to take into account the effect of the initial size. We checked on the mean-field model that this expression corrects at leading order the finiteness of $t_\textrm{TOF}$.
\end{thebibliography}
\end{document}